# Using current research information systems to investigate data acquisition and data sharing practices of computer scientists

Antti Mikael Rousi
Aalto University, Finland

## Abstract
Without sufficient information about research data practices occurring in a particular research organisation, there is a risk of mismatching research data service efforts with the needs of its researchers. This study describes how data acquiring and data sharing occurring within a particular research organisation can be investigated by using current research information system publication data. The case study organisation's current research information system was used to identify the sample of investigated articles. A sample of 193 journal articles published by researchers in the computer science department of the case study's university during 2019 were extracted for scrutiny from the current research information system. For these 193 articles, a classification of the main study types was developed to accommodate the multidisciplinary nature of the case department's research agenda. Furthermore, a coding framework was developed to capture the key elements of data acquiring and data sharing. The articles representing life sciences and computational research relatively frequently reused open data, whereas data acquisition of experimental research, human interaction studies and human intervention studies often relied on collecting original data. Data sharing also differed between the computationally intensive study types of life sciences and computational research and the study types relying on collection of original data. Research data were not available for reuse in only a minority of life science ($n = 2$; 7%) and computational research ($n = 15$; 14%) studies. The study types of experimental research, human interaction studies and human intervention studies less frequently made their data available for reuse. The findings suggest that research organisations representing computer sciences may include different subfields that have their own cultures of data sharing. This study demonstrates that analyses of publications listed in current research information systems provide detailed descriptions how the affiliated researchers acquire and share research data.

## Introduction

The key aims of open science – accessibility, reproducibility and transparency of science – are dependent on shared research data (Gutierrez et al., 2021; Tenopir et al., 2020). Academic libraries are supporting research data sharing and the findable, accessible, interoperable and reusable research data (FAIR data) movement (Cox et al., 2019; Huang et al., 2021; Wilkinson et al., 2016). For instance, the European Commission aims at making all research data generated in EU-funded projects as FAIR as possible (European Commission [EC], 2019). However, this can be difficult since various fields of science have different challenges when moving towards research data sharing (Borgman, 2012). In life sciences, data sharing in community-endorsed repositories is often expected (Rousi and Laakso, 2020), while in behavioural sciences, protecting the anonymity of the participants needs to be addressed before data can be shared (Taylor, 2017). Research organisations may employ researchers that represent subfields with different cultures of data sharing. Without precise information on how researchers working in a particular organisation acquire and share research data, there is a risk that research data service efforts will not match their needs.

Scientists' data sharing has often been examined as a part of multidisciplinary survey questionnaires (e.g. Kim

**Corresponding author:**
Antti Mikael Rousi, Research Services, Aalto University, Otakaari 1 A, Aalto 00076, Finland.
Email: antti.m.rousi@aalto.fi

and Zhang, 2015; Tenopir et al., 2011, 2020). Although these studies have been highly valuable, it is not clear how well the sample of scientists observed in multinational and multidisciplinary survey questionnaire studies reflect single research organisations, such as university departments. Institutional survey questionnaires (e.g. Chawinga and Zinn, 2020) more reliably gauge research data sharing at the level of individual research organisations. However, the response rates in institutional questionnaires may turn out so low that it is unclear whether all cultures of data sharing are sufficiently represented in the sample of responses. Data interviews (e.g. Abduldayan et al., 2021; Mallasvik and Martins, 2020: 589) have also been a common method for acquiring insight into researchers' data sharing. Although this methodology has also been valuable, inferring a holistic picture about a research organisation may be difficult based on sample interviews alone.

Scientific publications contain information about the underlying research data and previous research has investigated publications to gain a better understanding of research data sharing. Data availability statements have been investigated to assess the frequency of data sharing in articles of a particular journal (Federer et al., 2018). In addition, data sharing in biomedical research has been studied using random samples of articles (Wallach et al., 2018). Current research information systems (CRIS) are used to collect information about research organisations' scientific publications (Sivertsen, 2019; Velásquez-Duran and Ramírez Montoya, 2018). The current state-of-the-art CRIS also feature researcher profiles, integrated citation metrics and multimedia integrations (e.g. Elsevier, 2020). CRIS-listed publications of a particular research organisation could be analysed to investigate the research data practices of the affiliated researchers, and these analyses could complement the current questionnaire and interview methodologies eliciting insight into researchers' data sharing.

Previous studies of CRIS have focussed on their research-data related functionalities, such as the inclusion of data management plans, research data and research data metadata (Jetten et al., 2019; Schöpfel et al., 2017; Simons et al., 2017; see also Schöpfel et al., 2020). However, so far, the full potential of CRIS in acquiring insight into researchers' data sharing has not been realised. To elaborate these matters, this study describes how data acquiring and data sharing occurring in a particular research organisation can be investigated by analysing CRIS publication data. The present study was arranged around the following main research tasks. First, a sample set of publications were extracted from the case research organisation's CRIS. Second, a classification of the main study types exhibited in the sample set of articles was developed to accommodate the multidisciplinary nature of the case organisation's research agenda. Then, a coding framework was developed to capture the key elements of data acquiring and data sharing exhibited in the investigated articles. Lastly, how the types of data acquisition and data sharing fluctuated per the identified main study types were observed.

The target organisation of this case study was a department of computer science in a Finnish university. Prior studies suggest that computer scientists have positive attitudes about research data sharing (Tenopir et al., 2020: 15) and the issues of data management are being discussed within the field (e.g. Ivie and Thain, 2019; Stodden, 2020; Wilms et al., 2020). Furthermore, some of the subfields of computer sciences can be seen as data intensive (Maxim et al., 2021). These details make the field particularly interesting from the viewpoint of data sharing and FAIR data. Computer sciences may provide important examples and best practices for the FAIR data movement.

Given that the concept of research data may be elusive, the key element in the present study's approach to research data was reproducibility of science. If the value proposition of a journal article relied on data-based observations, research data was defined as the materials (in full or part) that allows readers to verify the authors' conclusions. The following terminology is at the core of this research. *Data acquisition* refers to activities used to obtain the research data. *Data reuse* is a specific form of data acquisition where openly available research data is reused to produce novel original research. *Data sharing* refers to efforts undertaken by the researchers to make their data available for reuse outside of their immediate research group or research project consortium. *Research data practices* is used as an umbrella term encompassing the above concepts. Figure 1 presents the core terminology of the present study.

## Literature review

### Multidisciplinary studies of scientists' data acquiring and sharing

As mentioned in the introduction, scientists' data sharing has often been examined as a part of multidisciplinary questionnaire studies (Joo and Kim, 2017; Kim, 2017; Kim and Zhang, 2015; Tenopir et al., 2011, 2020; Ünal et al., 2019; Wilms et al., 2020). Although these studies have been highly valuable in assessing the general state of data sharing within different fields of science., it is not clear how well the sample of, for example, computer scientists observed in multidisciplinary studies reflect single university departments. Institutional survey questionnaires (Chawinga and Zinn, 2020) more reliably gauge research data sharing at the level of individual research organisations. However, the response rate in institutional questionnaires may turn out so low that it is unclear whether all cultures of data sharing within, for example, particular university departments are sufficiently represented in the sample of responses.

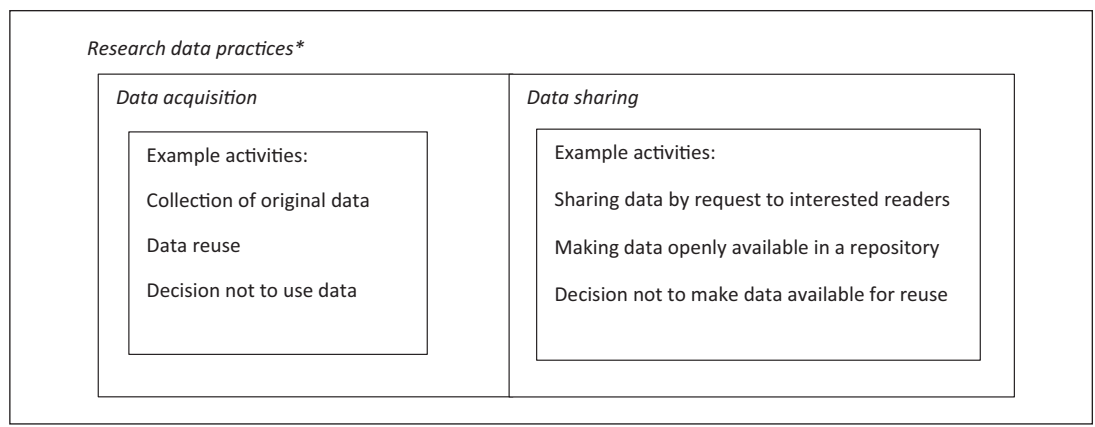

**Figure 1.** The core concepts of the present study. * = If the value proposition of a journal article relies on data-based observations, research data is defined as the materials (in full or part) that allows readers to verify the authors' conclusions.

Data interviews (e.g. Abduldayan et al., 2021; Mallasvik and Martins, 2020; Poole and Garwood, 2020; Suhr et al., 2020) have also been a common method for acquiring insight into researchers' data sharing. Although the interview methodology has also been valuable in understanding, for example, researchers' motivations and barriers to data sharing, inferring a holistic picture about a research organisation may be difficult based on sample interviews alone. Furthermore, the local nature of research data management services (e.g. Cox et al., 2019; Huang et al., 2021) and varying national legislation (e.g. Wilms et al., 2020: 8) highlight the importance of reliably examining data sharing at the level of individual research organisations.

### Studies of computer scientists' research data acquiring and sharing

Computer scientists' data sharing has often been examined as a part of multidisciplinary questionnaire studies (Joo and Kim, 2017; Kim, 2017; Kim and Zhang, 2015; Tenopir et al., 2011, 2020; Ünal et al., 2019; Wilms et al., 2020). The questionnaire studies suggest that computer scientists have positive thoughts about data sharing and that the reuse of open data is somewhat frequent within the field (Tenopir et al, 2011; Tenopir et al., 2020: 15). One of the recurring themes of these studies is that assistance (e.g. information about the basic rights of researchers) and infrastructures (e.g. data repositories) are important prerequisites for data sharing (Joo and Kim, 2017: 881; Wilms et al., 2020: 8; also Chawinga and Zinn, 2019: 116–118). Interview studies of computer scientists are scarce. However, insight to computer scientists' data sharing may be drawn from studies examining engineering research in general. Interview studies conducted within the fields of engineering research suggests that the institutional contexts are important for data sharing. The findings suggest that institutional rules, services and attitudes mediate the engagement in data sharing practices within engineering research (Mallasvik and Martins, 2020: 589; Abduldayan et al., 2021).

### Studies of CRIS

Previous studies of CRIS have focussed on their research data related functionalities, such as the inclusion of data management plans, research data and metadata in the systems (Jetten et al., 2019; Schöpfel et al., 2017; Simons et al., 2017; see also Schöpfel et al., 2020). The potential role of CRIS systems as a data source for science of science research has also been acknowledged (Sivertsen, 2019). However, so far, the full potential of CRIS in acquiring insight into researchers' data sharing has not been realised. To elaborate these matters, this study describes how data acquiring and data sharing occurring in a particular research organisation can be investigated by analysing CRIS publication data.

### Other related studies

As alluded to in the introduction, previous research has investigated scientific publications to gain a better understanding of research data sharing. Prior studies have analysed scientific publications to investigate topics such as the citation advantage linked to data sharing (Colavizza et al., 2020) and how subsequent publications reuse the data of highly cited articles (Imker et al., 2021). Data availability statements have been investigated to assess the frequency of data sharing in a particular journal (Federer et al., 2018). In addition, data sharing in biomedical literature has been studied using random samples of journal articles (Wallach et al., 2018). The related studies also include research on data management planning (Gajbe et al., 2021; Palsdottir, 2021), the role academic libraries in providing data management services (Cox et al., 2017, 2019; Huang et al., 2021), and research data metadata quality (Kim and Burns, 2016; Quarati and Raffaghelli, 2020).

Although prior research has been valuable, inferring a holistic picture about the data practices occurring in particular a research organisation may be difficult based on questionnaire and interview studies alone. Without sufficient information about the current practices, there is a risk of mismatching research data service efforts with the needs of the researchers. Therefore, the advancement of open and FAIR data could benefit from novel CRIS-based methods that reliably gauge data acquiring and sharing at the level of individual research organisations.

## Methodology

### Research aim

The aim of this study is to examine how research data is acquired and shared in the scientific articles authored by scientists affiliated with the computer science department of the case study's university. The case study organisation's CRIS was used to identify the sample of investigated articles. The main research questions are posed as follows:

> RQ1: What are the main study types that can be identified from the sample set of journal articles?
> RQ2: What types of data acquisition and data sharing can be identified from the sample set of journal articles?
> RQ3: How do the types of data acquisition and data sharing fluctuate per the main study types within the sample of articles?

### Case study organisation

The target organisation of this case study was a department of computer science in a Finnish university. The case department currently employs 44 professors, c. 70 post-docs and c. 100 PhD students, totalling over 450 employees. In 2019, the department was ranked within top 10 European computer science departments in the ShanghaiRanking's Global Ranking of Academic Subjects, Computer Science & Engineering (ShanghaiRanking, 2021).

### Data collection

The sampling strategy was focussed on capturing the key elements of data acquiring and data sharing exhibited in the scientific articles published by the computer scientists affiliated to the case organisation. Although prior studies have shown that engineers and scientists use diverse channels in their scientific communication (Wellings and Casselden, 2019: 795; see also Athukorala et al., 2013), the present study was limited to journal articles, as these often represent the final research output for many of the research groups of the case department. Prior studies show that researchers are reluctant to share data before finalising their own publication process (Tenopir et al., 2020: 18). Therefore, data sharing during pre-publication activities such as conference presentations or proceedings was not examined in the present study.

The CRIS was used to identify all scientific articles published by authors affiliated with the computer science department of the case university during the year 2019. The first set of extracted data included a total of 201 scientific articles, which included at least one author from the case computer science department. Non-peer reviewed editorials (e.g. special issue introductions) were screened from the sample as they did not represent original research. Furthermore, the analysis was based on both full-texts and metadata of the articles; individual works were excluded if the full-text was not accessible. The final data comprised 193 articles. Figure 2 depicts the flow of information in the sampling procedure using the PRISMA guidelines for systematic literature reviews (Moher et al., 2009).

### Data analysis

Already in the early phases of research it was understood that it was necessary to perform the analyses of the articles' main study types and data acquisition and data sharing manually from the article full texts and the metadata provided in the publishers' webservices. As the national field of science metadata classification works at a general level (publications are classified to represent, for example, computer sciences or physical sciences), the CRIS metadata could not be used to infer the work's main study types with sufficient precision. Furthermore, although the CRIS of the case study's university allow the inclusion of metadata for research data, this feature could not be used to provide accurate information on how individual publications acquired and shared research data. Inclusion of research data metadata was a relatively new feature in the CRIS of the case study's university, and researchers only rarely added this information during 2019.

The journal articles' full texts and metadata located in the publishers' services were scrutinised first for the main study type and then for data acquisition and data sharing. A classification framework was applied to categorise the main study types and research data acquisition and sharing used in the investigated articles. The classification framework was developed through qualitative analysis of the previously described set of computer science journal articles by means of a constant comparative method (Silverman, 2005). Preliminary coding was first done to incorporate the above elements, and then further enhanced through comprehensive data treatment until no new variants could be inferred from the data. Once the final classification framework was formed, the data was scrutinised again to ensure the consistency of the findings.

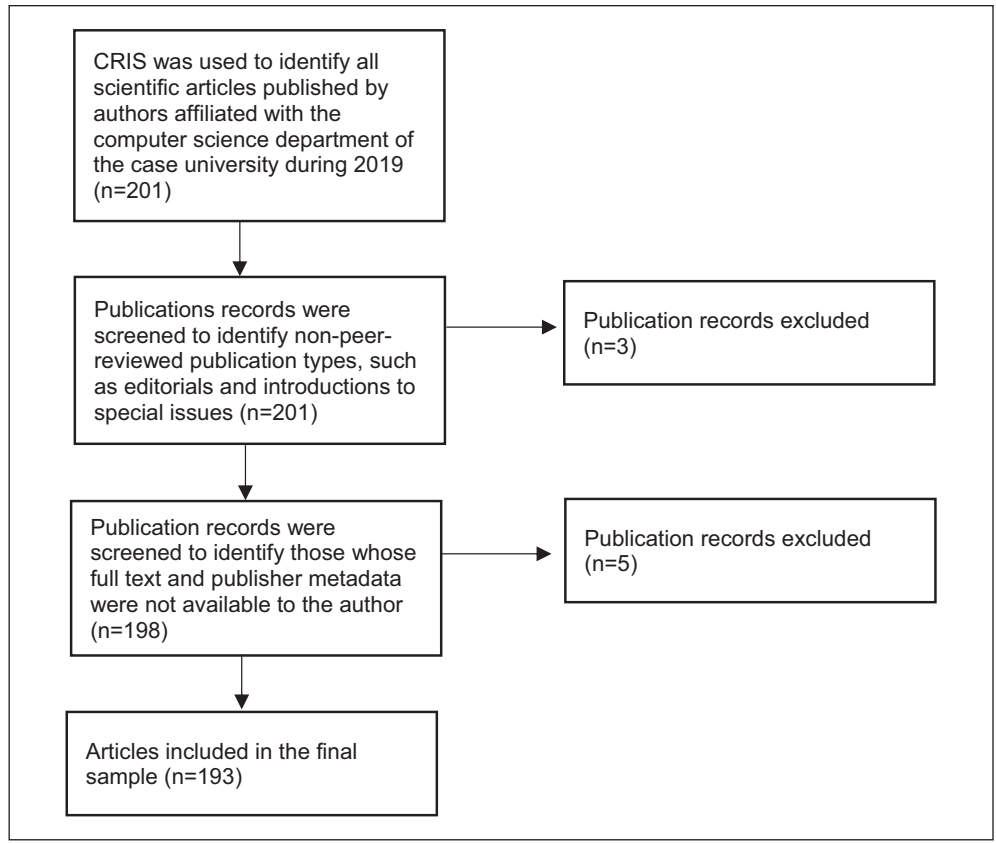

**Figure 2.** The flow of information in the sampling procedure depicted using the PRISMA (Moher et al., 2009) guidelines.

As mentioned in the introduction, the key element in the approach to research data was reproducibility of science. If the value proposition of the journal article relied on data-based observations, research data was defined as the materials (in full or part) that allowed readers to verify the authors' conclusions. Within the present study, research data could thus take the forms of genomic sequence data, questionnaire data collected from human participants, and ecological data. Artificial or randomly generated training data (also synthetic data) were also treated as original data. Descriptions of variables or data collection methods or additional analyses provided as supplementary materials were not considered as research data in the present study. Similarly, simulation parameters (novel or from prior studies), as well as Monte Carlo simulations that were explained in the article texts were not treated as research data.

### Classification framework

To draw attention to the multidisciplinary nature of the case computer science department's research agenda, a rudimentary classification of the main study types exhibited in the investigated articles was developed. This classification was primarily developed through qualitative analysis of the investigated articles, but it was also informed by the laws and regulations affecting the case organisation. Given the strict European data protection regulation (GDPR; see European Union [EU], 2020), it seen as important to identify the share of studies that collect data from human subjects, for example. This analysis identified the following six main study types:

- *Computational research* – a set of studies where the main value proposition of the article comes from presenting a novel computational method, for example articles reporting algorithms, machine learning solutions, simulations and software.
- *Experimental research* – a set of studies where the main value proposition comes from novel measurement results from different experimental settings, for example studies of astronomy, meteorology, acoustics and chemistry.
- *Interaction studies* – studies where data is collected from human participants through ways of interaction, such as studies using interviews, questionnaires or ethnography as their method of data collection. Within the context of the case organisation, studies representing this study type are often regulated by the European general data protection regulation (EU, 2020).

**Table 1.** Classification framework used in the present study.

| Data points | Coding |
|---|---|
| Main study types | |
|     Computational research (e.g. algorithms, simulations, software) | yes/no |
|     Experimental research (e.g. astronomy, weather, acoustics, chemistry) | yes/no |
|     Interaction studies (e.g. interviews, questionnaires, tasks, ethnography) | yes/no |
|     Intervention studies (e.g. EEG, MRI, clinical trials) | yes/no |
|     Life sciences (e.g. genomics, bioinformatics, ecology) | yes/no |
|     Literature or conceptual studies (e.g. literature reviews, archive studies, social media studies) | yes/no |
| Data acquisition | |
|     Authors presented a data availability statement | yes/no |
|     Authors did not utilise data | yes/no |
|     Authors collected original data | yes/no |
|     Authors reused open data from prior studies | yes/no |
|     Authors reused open data from public authorities, companies, universities or associations | yes/no |
| Data sharing | |
|     Authors did not utilise data | yes/no |
|     Data of the article were not available for reuse | yes/no |
|     Authors used openly available data | yes/no |
|     Authors agreed to share their data to interested readers | yes/no |
|     Authors shared data (or part of it) as supplementary material | yes/no |
|     Authors shared data (or part of it) via open deposition | yes/no |
|     Authors deposited code or used open code | yes/no |

The categories presented in main study types, data acquisition and data sharing are not mutually exclusive.

- *Intervention studies* – studies where data is collected from human or animal participants through *in vivo* intervention methods. This main study type includes, but is not limited to, medical research. Study types such as clinical trials, magnetic resonance imaging (MRE), and electroencephalography (EEG) are included in this category. Within the context of the case organisation, these types of studies are often regulated by the national law on medical research (Ministry of Social Affairs and Health, 1999), national ethical guidelines of human sciences (Finnish National Board on Research Integrity, 2019), and the EU's GDPR (EU, 2020).
- *Life sciences* – studies that examine so-called 'omics', such as genomics or proteomics, as well as studies of biology and ecology.
- *Literature or conceptual studies* – a meta study or review of prior scientific articles or other documents, for example studies examining primarily prior scientific literature, newspaper or patent archives or social media archives.

In addition to these characterisations of main study types, classifications of data acquisition and data sharing were developed through the qualitative analysis of the articles as data. Table 1 presents the final classification framework used in the present study. The categories presented in main study types, data acquisition and data sharing are mutually non-exclusive, for example an article could have several main study types and also use data from prior articles and from public databases.

With research data, it was often easily to distinguish whether a specific set of data was generated as a result of the experiment explained in the article. However, with open code, the distinction of whether the used code was generated for the purposes of a specific article or whether the code was deposited earlier and used in prior studies was often not clear. Therefore, code sharing was examined at a more general level, that is, whether an 'Authors deposited code or used open code' (see Table 1).

Regarding code sharing, only explicit references to codes and software were counted as use of open code. The mere mention of code, for example 'gym Pendulum-v0 environment' without any link or reference was not interpreted as use of open code. It is noteworthy that even though not all studies of novel computational methods deposited source code, this does not mean that they were not reproducible. Within studies examining, for example, novel algorithms, the details of the algorithm were often described in the article's methodology section.

The collected data are provided as open data to facilitate future research (Rousi, 2021). The data are archived and described here https://doi.org/10.5281/zenodo.4736881

## Findings

### Data acquisition

The computer science articles investigated ($n=193$) both collected original data and re-used openly available data from a variety of sources. Table 2 presents how data

**Table 2.** Data acquisition per main study type.

| Main study types | Authors presented a data availability statement (%) | Authors did not utilise data (%) | Authors collected original data (%) | Authors reused open data from prior studies (%) | Authors reused open data from public authorities, companies, universities or associations (%) |
|---|---|---|---|---|---|
| Computational research (n = 111) | 18 (16) | 34 (31) | 36 (32) | 31 (28) | 37 (33) |
| Experimental research (n = 21) | 3 (14) | 0 | 18 (86) | 5 (24) | 4 (19) |
| Interaction studies (n = 40) | 5 (13) | 0 | 38 (95) | 3 (8) | 3 (8) |
| Intervention studies (n = 16) | 5 (31) | 0 | 14 (88) | 2 (13) | 2 (13) |
| Life sciences (n = 27) | 13 (48) | 1 (4) | 12 (44) | 12 (44) | 16 (59) |
| Literature or conceptual studies (n = 17) | 1 (6) | 5 (29) | 11 (65) | 0 | 1 (6) |
| All examined articles (n = 193) | 30 (16) | 38 (20) | 106 (55) | 42 (22) | 45 (23) |

Number of articles (percentage of articles representing this study type). Categories are not mutually exclusive.

acquisition varied within the investigated sample of articles. Original data were gathered in a total of 106 (55%) articles (Table 2). A total of 42 (22%) articles reused data from prior studies and a total of 45 (23%) reused open data provided by public authorities, companies, universities or associations. Notably, one fifth of all investigated articles (n = 38; 20%) did not use research data as a part of their value proposition. Data availability statements were rare, and only 30 (16%) of the investigated articles presented one.

There were significant differences in the ways in which authors of articles representing different types of research acquired data. Experimental research and human interaction and intervention studies relatively frequently collected original data. Eighteen (86%) experimental research studies, 38 (95%) interaction studies and 14 (88%) intervention studies collected original data (Table 2). Research data reuse was relatively frequent within articles representing life sciences and computational research. Twelve (44%) life science studies and 31 (28%) computational research studies reused data from prior studies. Notably, the study type of computational research had the highest share of articles that did not use research data (n = 34; 31%).

Within the articles investigated, open research data were acquired from a variety of sources. Table 3 presents the public authorities, companies, universities and associations from which open research data were acquired. Especially computational research and life science studies reused openly available data from diverse sources (see Table 3). Besides established data repositories, computational research studies sometimes used open data posted on researchers' websites.

### Data sharing

Data sharing was frequent within the investigated sample of articles. A total of 155 of the investigated computer science articles acquired research data from which a total of 92 (59%) articles discussed some level of openly available data. Table 4 presents how the different types of data sharing fluctuated within the sample of articles. The data of 25 (13%) articles were shared as supplementary materials accompanying the published articles, and the data of 30 (16%) articles were deposited to a repository (see Table 4). Within a small minority of articles (n = 7; 4%), it was suggested that the authors are willing to share the data upon a request. The use of open code was somewhat common, as a total of 79 (41%) articles either deposited or used open code.

Although data sharing occurred in all of the main study types, there were significant differences in the frequency of sharing (see Table 4). Life science studies relatively frequently made data available for reuse. Ten (37%) of the life science studies shared data through supplementary material and 12 (44%) of life science studies shared data through repositories. On the other hand, human interaction studies used open data or shared data the least: 30 (75%) interaction studies did not make data available for reuse (Table 4).

The use of open code also occurred in all of the main study types, but there again were differences in frequency of this use. The use of open code was most common within life science and computational research studies. Twenty-one (78%) life science studies and 63 (57%) computational studies either made code available for reuse or used openly available code (Table 4).

The authors of the investigated articles used a variety of repositories and services for data sharing. Table 5 presents the identified repositories and services used for data sharing. It seems noteworthy that although only 27 of the investigated articles were life science studies, these articles exhibited diverse services used for sharing data.

### Discussion

This study described how data acquiring and data sharing occurring within a particular research organisation can be investigated by using CRIS publication data. The case study organisation's CRIS was used to identify the sample

**Table 3.** Example sources of open data categorised per main study types.

| Main study types | Example sources of open data (number of articles) |
| --- | --- |
| Computational research* | UCI Machine Learning Repository (University of California) (*n* = 4) |
|  | Researcher website (*n* = 2) |
|  | American Physical Society (citation data) (*n* = 1) |
|  | Cambridge Structural Database (*n* = 1) |
|  | developer.here.com (*n* = 1) |
|  | Digital Bibliography and Library Project (DBLP) (*n* = 1) |
|  | Extreme Classification (XMC) Repository (*n* = 1) |
|  | Finnish Meteorological Institute (*n* = 1) |
|  | Github (*n* = 1) |
|  | Google Cluster data 2019 (*n* = 1) |
|  | Helsinki Region Transport (*n* = 1) |
|  | Kaggle (*n* = 1) |
|  | Music-IR (MIREX) (*n* = 1) |
|  | National Library of Finland (*n* = 1) |
|  | NeurIPS Proceedings (*n* = 1) |
|  | Quantum-machine.org (*n* = 1) |
|  | Research Group Website (*n* = 1) |
|  | ShapeNet (*n* = 1) |
|  | Stanford SNAP repository (*n* = 1) |
|  | Statistics Finland (*n* = 1) |
|  | SUNCG dataset (Princeton University) (*n* = 1) |
|  | Time Series Classification Archive (University of California) (*n* = 1) |
|  | Twitter (*n* = 1) |
|  | UC Irvine Network Data Repository (*n* = 1) |
|  | UC Merced Land Use Dataset (University of California) (*n* = 1) |
|  | US National Climatic Data Center (*n* = 1) |
|  | Web of Science (*n* = 1) |
|  | World bank (*n* = 1)) |
| Experimental research | Finnish National Meteorological Institute (*n* = 1) |
|  | Foreca Oy (*n* = 1) |
|  | Heliophysics Integrated Observatory (HELIO) (*n* = 1) |
|  | National Oceanic and Atmospheric Administration (NOAA) (*n* = 1) |
|  | Nordic Optical Telescope Scientific Association (*n* = 1) |
| Interaction studies | Digital Bibliography & Library Project (DBLP) (*n* = 1) |
|  | Foreca Oy (*n* = 1) |
|  | Pohtiva (A repository of Finnish political programmes) (*n* = 1) |
| Intervention studies | Agency for Healthcare Research and Quality (AHRQ) (*n* = 1) |
|  | The Genotype-Tissue Expression (GTEx) project (*n* = 1) |
| Life sciences* | Genomics of Drug Sensitivity in Cancer (GDSC) (*n* = 2) |
|  | Global Natural Products Social Molecular Networking (GNPS) (*n* = 2) |
|  | String (*n* = 2) |
|  | BiGG Models (University of California) (*n* = 1) |
|  | Cancer Cell Line Encyclopaedia (CCLE) (*n* = 1) |
|  | Cancer Therapeutics Response Portal (CTRP) (*n* = 1) |
|  | Centers for Disease Control and Prevention (CDC) (*n* = 1) |
|  | Genemania (*n* = 1) |
|  | Geneontology (*n* = 1) |
|  | GTEX portal (*n* = 1) |
|  | Human Phenotype Ontology (*n* = 1) |
|  | MassBank Europe (*n* = 1) |
|  | NCBI Gene expression omnibus (*n* = 1) |
|  | Reactome (*n* = 1) |
|  | Swissprot (*n* = 1) |
|  | The Global Pneumococcal Sequencing Project (*n* = 1) |
|  | The Cancer Genome Atlas (TCGA) (*n* = 1) |
|  | Victorian Biodiversity Atlas (*n* = 1) |
|  | Wellcome Sanger Institute (*n* = 1) |
| Literature or conceptual studies | Github (*n* = 1) |

*A total of 18 articles were classified to have both computational research and life sciences as their main study types. All data sources utilised in these previous articles that contained bioinformatics data are presented in the life sciences category.

**Table 4.** Data sharing per main study type.

| Main study types | Authors did not utilise data (%) | Data of the article were not available for reuse (%) | Authors used openly available data (%) | Authors agreed to share their data to interested readers (%) | Authors shared data (or part of it) as supplementary material (%) | Authors shared data (or part of it) via open deposition (%) | Authors deposited source code or used open code (%) |
|---|---|---|---|---|---|---|---|
| Computational research ($n=111$) | 34 (31%) | 15 (14) | 54 (49) | 3 (3) | 12 (11) | 14 (13) | 63 (57) |
| Experimental research ($n=21$) | 0 | 12 (57) | 7 (33) | 0 | 1 (5) | 4 (19) | 6 (29) |
| Interaction studies ($n=40$) | 0 | 30 (75) | 3 (8) | 3 (8) | 3 (8) | 6 (15) | 6 (15) |
| Intervention studies ($n=16$) | 0 | 8 (50) | 3 (19) | 2 (13) | 5 (31) | 4 (25) | 5 (31) |
| Life sciences ($n=27$) | 1 (4%) | 2 (7) | 17 (63) | 2 (7) | 10 (37) | 12 (44) | 21 (78) |
| Literature or conceptual studies ($n=17$) | 5 (29%) | 6 (35) | 1 (6) | 0 | 4 (24) | 1 (6) | 2 (12) |
| All examined articles ($n=193$) | 38 (20%) | 63 (33) | 65 (34) | 7 (4) | 25 (13) | 30 (16) | 79 (41) |

Number of articles (percentage of articles representing this study type). Categories are not mutually exclusive.

**Table 5.** Example services used for data sharing as categorised per main study types.

| Main study types | Example services used for data sharing (number of articles) |
|---|---|
| Computational research* | Data Archive for Social Sciences (GESIS) ($n=1$) |
| | GitHub ($n=3$) |
| | Journal's archive ($n=1$) |
| | Research Group Website ($n=1$) |
| | Zenodo Repository ($n=1$) |
| Experimental research | Research Group Website ($n=1$) |
| | Research Project Website ($n=1$) |
| | Strasbourg astronomical Data Center (CDS) ($n=2$) |
| Interaction studies | Dryad Digital Repository ($n=1$) |
| | Finnish National Institute for Health and Welfare ($n=1$) |
| | Open Science Foundation (osf.io) ($n=1$) |
| | Project Website ($n=2$) |
| | Zenodo Repository ($n=1$) |
| Intervention studies | European Genome-phenome Archive ($n=1$) |
| | GitHub ($n=1$) |
| | NeuroVault ($n=1$) |
| | Open Science Foundation (osf.io) ($n=1$) |
| Life sciences* | Github ($n=4$) |
| | European Genome-phenome Archive (EGA) ($n=2$) |
| | Gene Expression Omnibus (GEO) ($n=2$) |
| | Research group's website ($n=2$) |
| | ArrayExpress ($n=1$) |
| | AstraZeneca Open Innovation Portal ($n=1$) |
| | Catalogue of Somatic Mutations in Cancer (COSMIC) ($n=1$) |
| | Dryad Digital Repository ($n=1$) |
| | European Nucleotide Archive ($n=1$) |
| | Genomics of Drug Sensitivity in Cancer (GDSC) ($n=1$) |
| | Global Pneumococcal Sequencing Project (GPS) ($n=1$) |
| | MassIVE (University of California) ($n=1$) |
| | NCBI Sequence Read Archive ($n=1$) |
| | NCBI GenBank ($n=1$) |
| | Synapse database ($n=1$) |
| Literature or conceptual studies | Zenodo Repository ($n=1$) |

*A total of 18 articles were classified to have both computational research and life sciences as their main study types. All data repositories utilised in these previous articles that contained bioinformatics data are presented in the life sciences category.

of investigated articles representing computer sciences ($n$ = 193). In general, the investigated articles both collected original data, re-used openly available data, and frequently made data available for reuse. However, the types and frequency of data acquisition and data sharing differed between the identified main study types. Thus, the findings suggest that research organisations representing computer sciences are not monoliths in their data practices and may include different subfields that have their own cultures of data sharing.

The articles representing life sciences and computational research relatively frequently reused open data, whereas data acquisition of experimental research, human interaction studies and human intervention studies often relied on collection of original data (see Table 2). Data sharing also differed between these computationally intensive study types of computational research and life sciences and the study types relying on original data. Research data was not available for reuse in only a minority of life science ($n$ = 2; 7%) and computational research studies ($n$ = 15; 14%). Open code was also relatively frequent within these computationally intensive types of research. Twenty-one (78%) of life science studies and 63 (57%) of computational research studies either reused or shared open code. The study types of experimental research, human interaction studies and human intervention studies – that often relied on collection original data – less frequently made their data or code available for reuse (see Table 4). These differences may be at least partly explained by both the regulations affecting individual subfields (e.g. GDPR regulation and human-interaction and intervention studies within the context of Europe, see EU, 2020) and the state of the established methods of data sharing within the different subfields (e.g. the use of community-endorsed repositories within life sciences).

When compared to Wallach et al.'s (2018) study of biomedical journal articles, data sharing was more frequent within the computer science articles investigated in this study. The findings seem to support previous research indicating that computer scientists have relatively positive attitudes about research data sharing (Tenopir et al., 2020: 15). However, it is important to note that the present article reported a case study and the findings are not necessarily generalisable to all research organisations representing computer sciences. In addition, it seems important to note that in Federer et al.'s (2018) study of *PLOS One* articles, the share of articles that discussed openly available data was higher than in the present study. However, it is unclear how the prevalent data sharing categories of 'in paper' and 'in paper and SI [supplementary information]' of the Federer et al.'s (2018: 5) study correspond with the categorisation used in this study.

The findings suggest that analyses of CRIS publications can be used to complement questionnaire and interview methods eliciting insight into researchers' data sharing. Although institutional survey questionnaires (e.g. Chawinga and Zinn, 2020) do gauge data sharing at the level of individual research organisations, the response rate in these questionnaires may turn out so low that it is unclear whether all cultures of data sharing are sufficiently represented in the sample of responses. Whereas the interview studies are highly valuable in understanding, for example, researchers' motivations and barriers to data sharing, inferring a holistic picture about a particular research organisation with, for example, circa 45 research groups may be difficult based on sample interviews alone. As exemplified in this article, thorough analysis of research organisation's CRIS publications highlights the different subcultures of data practices and reveals the repositories used for data sharing.

The CRIS-based research design used here seems to hold a good promise to improve the understanding of how research data is shared within research organisations. In general, scientific articles seem to be a good source of information about data sharing occurring in research organisations. However, whether the findings that are based on a single year's publication activity predict this behaviour over longer timespans, for example over the next following five years, is unclear. This important question should be examined through longitudinal research designs. However, prior literature (e.g. Tenopir et al., 2015; van Panhuis et al., 2014: 18) suggests the cultures of data sharing can be slow to change. Thus, it seems reasonable to consider the findings as informative and use them to assist in the allocation of research data management services (e.g. Cox et al., 2019; Huang et al., 2021; Mons, 2018).

Given that CRIS systems have advanced functionality for data reporting, and that manual analysis of articles is a time-consuming task, a logical next step would be to either make the information regarding data acquisition and sharing in a machine-readable format, or use machine-learning solutions to mine this information from CRIS data (Azeroual et al., 2018; Biesenbender et al., 2019). However, it is important to note that the classification framework used in the present study was developed through a case study, hence the categories of main study types and research data acquisition and sharing may not be generalisable. More studies that elaborate the classification of research data sharing would be beneficial. A key question is whether these kinds of classifications should be international, domain-specific classifications or whether they should be approached nationally, or even at the level of individual research organisations, especially since the ethical regulation of human sciences may differ from country to country (e.g. U.S. Department of Health and Human Services, 2020). The ethics of such automated data processing should also be considered (Schöpfel et al., 2020).

The analysis highlighted the importance of giving detailed references both to data sets and open code. As discussed in the methodology section, the relation between data sets and articles was often easily distinguishable (i.e. data set X was part of the research that led to article Y; when citing data set X, I also cite the article Y). However, with references to open code, this relation was often not clear and the research from which the code originated was at times not mentioned. Furthermore, references to open code were often provided as footnotes instead of including them in the list of references. Further studies providing more detailed examinations of how data and code are being cited within computer sciences would be beneficial.

### Empirical, theoretical and practical implications

The contribution of the present study is foremost methodological in its nature. This study demonstrates that analyses of publications listed in CRIS provide detailed descriptions how the affiliated researchers acquire and share research data. As CRIS are often managed by academic libraries (e.g. Jetten et al., 2019), the presented research design may be adopted and developed by LIS scholars and professionals.

The practical implications of this work are related to how research data management services could be coordinated in the target organisation of this case study. Computational research was prevalent within the investigated articles. Thus, there could be a demand for research data management services – and code management in particular – within this line of research. The study types of human interaction, human intervention and experimental research all frequently collected original data, but this data was only rarely provided for reuse. Due to the strict privacy regulation (EU, 2020), the services may also need to launch and support initiatives that aim for facilitated sharing of data collected from human subjects. These initiatives may include providing services for anonymisation of personal data, services consent-based sharing of personal data (see e.g. Bannier et al., 2021) or utilising combinations of different technological solutions to promote data sharing (Durrant et al., 2021).

### Limitations

The main limitation of the present study is that it was a case study of a specific department within a specific university. Therefore, the classification framework and findings are not necessarily generalisable. Further studies of computer scientists' data sharing should be conducted to verify the generalisability of the results.

The present study also has a specific limitation, which could be described as the problem of relevant data sharing. The analysis of data sharing was done at a general level, and its key element was a general distinction whether the investigated articles shared research data or parts of it. Hence, whether the shared materials included all relevant data and necessary protocols to completely reproduce the results of the investigated scientific articles was not explored in detail (this was not a replication study). Given the multidisciplinary nature of the research of the case organisation, a thorough review regarding the reproducibility of individual articles would have required substantial expertise from diverse subfields of computer science. However, the findings may be still seen as informative, as they provide an overview regarding the frequency and methods of data sharing occurring in the different computer science subfields.

## Conclusions

This study described how research data acquiring and sharing occurring within a particular research organisation can be investigated by using CRIS publication data. The findings demonstrate that analyses of publications listed in CRIS provide detailed descriptions how the affiliated researchers acquire and share research data. Furthermore, the findings suggest that research organisations, such as university departments, may include subfields that have their own cultures of data sharing. Manual analysis of scientific articles is a time-consuming task, and automated processing, such as machine-readable metadata and machine learning solutions, could be used to extract research data-related information from CRIS. I suggest further qualitative studies to produce more detailed descriptions and classifications of data acquiring and sharing occurring in different scientific domains. Once a sufficient knowledge base is achieved, this insight could then be applied to CRIS data using different automated processing methods.

### Acknowledgements

The author would like to thank the anonymous reviewers for their comments that greatly helped to improve the manuscript. Dr. Jacquelin De Faveri helped in the English-language editing of the submitted manuscript.

### Authors' note

The pre-print version of this manuscript has been published in the arXiv Digital libraries collection: [2106.09399] Data sharing of computer scientists: an analysis of current research information system data (arxiv.org).

### Data availability statement

The collected data and coding framework are provided as open data to facilitate future research. The data are archived and described here https://doi.org/10.5281/zenodo.4736881

### Declaration of conflicting interests

The author declared no potential conflicts of interest with respect to the research, authorship, and/or publication of this article.

## Funding

The author(s) disclosed receipt of the following financial support for the research, authorship, and/or publication of this article: This study was supported by Aalto university

## ORCID iD

Antti Mikael Rousi  https://orcid.org/0000-0002-4184-7035

## References

Abduldayan FJ, Abifarin FP, Oyedum GU, et al. (2021) Research data management practices of chemistry researchers in federal universities of technology in Nigeria. *Digital Library Perspectives* 37(4): 328–348.

Athukorala K, Hoggan E, Lehtiö A, et al. (2013) Information-seeking behaviors of computer scientists: Challenges for electronic literature search tools. *Proceedings of the American Society for Information Science and Technology* 50(1): 1–11.

Azeroual O, Saake G, Abuosba M, et al. (2018) Text data mining and data quality management for research information systems in the context of open data and open science. Available at: https://arxiv.org/abs/1812.04298 (accessed 16 April 2022).

Bannier E, Barker G, Borghesani V, et al. (2021) The Open Brain Consent: Informing research participants and obtaining consent to share brain imaging data. *Human Brain Mapping*, 42(7): 1945–1951.

Biesenbender S, Petersohn S and Thiedig C (2019) Using Current Research Information Systems (CRIS) to showcase national and institutional research (potential): Research information systems in the context of Open Science. *Procedia Computer Science* 146: 142–155.

Borgman CL (2012) The conundrum of sharing research data. *Journal of the American Society for Information Science and Technology* 63(6): 1059–1078.

Chawinga WD and Zinn S (2019) Global perspectives of research data sharing: A systematic literature review. *Library & Information Science Research* 41(2): 109–122.

Chawinga WD and Zinn S (2020) Research data management at a public university in Malawi: The role of "three hands." *Library Management* 41(6/7): 467–485.

Colavizza G, Hrynaszkiewicz I, Staden I, et al. (2020) The citation advantage of linking publications to research data. *PLoS One* 15(4): e0230416.

Cox AM, Kennan MA, Lyon L, et al. (2017) Developments in research data management in academic libraries: Towards an understanding of research data service maturity. *Journal of the Association for Information Science and Technology* 68(9): 2182–2200.

Cox AM, Kennan MA, Lyon L, et al. (2019) Maturing research data services and the transformation of academic libraries. *Journal of Documentation* 75(6): 1432–1462.

Durrant A, Markovic M, Matthews D, et al. (2021) How might technology rise to the challenge of data sharing in agrifood? *Global Food Security* 28: 100493.

Elsevier (2020) Pure releases: highlights of the 5.19 release. Available at: https://www.elsevier.com/solutions/pure/roadmap/releases (accessed 18 June 2021).

European Commission (2019) The EU's open science policy. Available at: https://ec.europa.eu/info/research-and-innovation/strategy/goals-research-and-innovation-policy/open-science_en (accessed 18 June 2021).

European Union (2020) Data protection under GDPR. Available at: https://europa.eu/youreurope/business/dealing-with-customers/data-protection/data-protection-gdpr/index_en.htm (accessed 18 June 2021).

Federer LM, Belter CW, Joubert DJ, et al. (2018) Data sharing in PLOS ONE: An analysis of data availability statements. *PLoS One* 13(5): e0194768.

Finnish National Board on Research Integrity (2019) The ethical principles of research with human participants and ethical review in the human sciences in Finland. Available at: https://tenk.fi/en/ethical-review/ethical-review-human-sciences (accessed 18 June 2021).

Gajbe SB, Tiwari A and Singh RK (2021) Evaluation and analysis of Data Management Plan tools: A parametric approach. *Information Processing & Management* 58(3):1–17.

Gutierrez RR, Lefebvre A, Núñez-González F, et al. (2021) Towards adopting open and data-driven science practices in bed form dynamics research, and some steps to this end. *Earth Surface Processes and Landforms* 46(1): 47–54.

Huang Y, Cox AM and Sbaffi L (2021) Research data management policy and practice in Chinese university libraries. *Journal of the Association for Information Science and Technology* 72(4): 493–506.

Imker HJ, Luong H, Mischo WH, et al. (2021) An examination of data reuse practices within highly cited articles of faculty at a research university. *The Journal of Academic Librarianship* 47(4): 102369.

Ivie P and Thain D (2019) Reproducibility in scientific computing. *ACM Computing Surveys* 51(3): 1–36.

Jetten M, Simons E and Rijnders J (2019) The role of CRIS's in the research life cycle. A case study on implementing a FAIR RDM policy at Radboud University, the Netherlands. *Procedia Computer Science* 146: 156–165.

Joo YK and Kim Y (2017) Engineering researchers' data reuse behaviours: A structural equation modelling approach. *The Electronic Library* 35(6): 1141–1161.

Kim Y (2017) Fostering scientists' data sharing behaviors via data repositories, journal supplements, and personal communication methods. *Information Processing & Management* 53(4): 871–885.

Kim Y and Burns CS (2016) Norms of data sharing in biological sciences: The roles of metadata, data repository, and journal and funding requirements. *Journal of Information Science* 42(2): 230–245.

Kim Y and Zhang P (2015) Understanding data sharing behaviors of STEM researchers: The roles of attitudes, norms, and data repositories. *Library & Information Science Research* 37(3): 189–200.

Mallasvik ML and Martins JT (2020) Research data sharing behaviour of engineering researchers in Norway and the UK: uncovering the double face of Janus. *Journal of Documentation* 77(2): 576–593.

Maxim BR, Galster M, Mistrik I, et al. (2021) Data-intensive systems, knowledge management, and software engineering.

In: Mistrik I, Galster M, Maxim BR, et al. (eds) *Knowledge Management in the Development of Data-Intensive Systems*. Boca Raton, FL: CRC Press, pp.1–40.

Ministry of Social Affairs and Health (1999) Finnish Medical Research Act. Available at: https://www.finlex.fi/fi/laki/kaannokset/1999/en19990488.pdf (accessed 18 June 2021).

Moher D, Liberati A, Tetzlaff J, et al. (2009) Preferred reporting items for systematic reviews and meta-analyses: The PRISMA statement. *Annals of Internal Medicine* 151(4): 264–269, W64.

Mons B (2018) *Data Stewardship for Open Science: Implementing FAIR Principles*. Boca Raton, FL: CRC Press.

Palsdottir A (2021) Data literacy and management of research data – A prerequisite for the sharing of research data. *Aslib Journal of Information Management* 73(2): 322–341.

Poole AH and Garwood DA (2020) Digging into data management in public-funded, international research in digital humanities. *Journal of the Association for Information Science and Technology* 71(1): 84–97.

Quarati A and Raffaghelli JE (2020) Do researchers use open research data? Exploring the relationships between usage trends and metadata quality across scientific disciplines from the Figshare case. *Journal of Information Science*. Epub ahead of print 4 October. DOI: 10.1177/0165551520961048.

Rousi AM (2021) Data of "Using current research information systems to investigate data acquisition and data sharing practices of computer scientists" [Data set]. *zenodo*. DOI: 10.5281/zenodo.4736881.

Rousi AM and Laakso M (2020) Journal research data sharing policies: A study of highly-cited journals in neuroscience, physics, and operations research. *Scientometrics* 124(1): 131–152.

Schöpfel J, Azeroual O and Jungbauer-Gans M (2020) Research ethics, open science and CRIS. *Publications* 8(4): 51.

Schöpfel J, Prost H and Rebouillat V (2017) Research data in current research information systems. *Procedia Computer Science* 106: 305–320.

ShanghaiRanking (2021) ShanghaiRanking's global ranking of academic subjects 2019, computer science & engineering. Available at: https://www.shanghairanking.com/ (accessed 18 June 2021).

Silverman D (2005) *Doing Qualitative Research*, 2nd edn. London: SAGE.

Simons E, Jetten M, Messelink M, et al. (2017) The important role of CRIS's for registering and archiving research data: The RDS-project at Radboud University (the Netherlands) in cooperation with data-archive DANS. *Procedia Computer Science* 106: 321–328.

Sivertsen G (2019) Developing Current Research Information Systems (CRIS) as data sources for studies of research. In: Glänzel W, Moed HF, Schmoch U, et al. (eds) *Springer Handbook of Science and Technology Indicators*. Cham, Switzerland: Springer, pp.667–683.

Stodden V (2020) The data science life cycle: A disciplined approach to advancing data science as a science. *Communications of the ACM* 63(7): 58–66.

Suhr B, Dungl J and Stocker A (2020) Search, reuse and sharing of research data in materials science and engineering—A qualitative interview study. *PLoS One* 15(9): e0239216.

Taylor L (2017) Safety in numbers? Group privacy and big data analytics in the developing world. In: Taylor L, Floridi L and van der Sloots B (eds) *Group Privacy: New Challenges to Data Technologies*. Cham, Switzerland: Springer, pp.13–36.

Tenopir C, Allard S, Douglass K, et al. (2011) Data sharing by scientists: Practices and Perceptions. *PLoS One* 6(6): e21101.

Tenopir C, Dalton ED, Allard S, et al. (2015) Changes in data sharing and data reuse practices and perceptions among scientists worldwide. *PLoS One* 10(8): e0134826.

Tenopir C, Rice NM, Allard S, et al. (2020) Data sharing, management, use, and reuse: Practices and perceptions of scientists worldwide. *PLoS One* 15(3): e0229003.

Ünal Y, Chowdhury G, Kurbanoğlu K, et al. (2019) Research data management and data sharing behaviour of university researchers. *Information Research* 24(1): 1–23.

U.S. Department of Health and Human Services (2020) International compilation of human research standards. Available at: https://www.hhs.gov/ohrp/international/compilation-human-research-standards/index.html (accessed 18 June 2021).

van Panhuis WG, Paul P, Emerson C, et al. (2014) A systematic review of barriers to data sharing in public health. *BMC Public Health* 14(1): 1144–1149.

Velásquez-Duran A and Ramírez Montoya MS (2018) Research management systems: systematic mapping of literature (2007-2017). *International Journal on Advanced Science Engineering and Information Technology* 8(1): 44–55.

Wallach JD, Boyack KW and Ioannidis JPA (2018) Reproducible research practices, transparency, and open access data in the biomedical literature, 2015-2017. *PLoS Biology* 16(11): e2006930.

Wellings S and Casselden B (2019) An exploration into the information-seeking behaviours of engineers and scientists. *Journal of Librarianship and Information Science* 51(3): 789–800.

Wilkinson MD, Dumontier M, Aalbersberg IJ, et al. (2016) The FAIR Guiding Principles for scientific data management and stewardship. *Scientific Data* 3: 160018.

Wilms KL, Stieglitz S, Ross B, et al. (2020) A value-based perspective on supporting and hindering factors for research data management. *International Journal of Information Management* 54: 102174.

## Author biography

Antti Mikael Rousi is a member of the Aalto University's Research Services. He has worked extensively with CRIS systems and open science. He holds a PhD in Information Science.